\documentclass[12pt,twoside]{article}

\usepackage{amssymb, amsfonts, amsmath}
\usepackage{mathrsfs,bm,color}

\catcode`\@=11

\def\newsymbol#1#2#3#4#5{\let\next@\relax%
 \ifnum#2=\@nehse%
 \ifnum#2=\tw@\let\next@\msyfam@\fi\fi%
 \mathchardef#1="#3\next@#4#5}
\def\mathhexbox@#1#2#3{\relax%
 \ifmmode\mathpalette{}{\m@th\mnnathchar"#1#2#3}
 hse\leavevmode\hbox{$\m@th\mathchar"#1#2#3$}\fi}
\font\tenmsy=msbm10
\font\sevenmsy=msbm7
\font\fivemsy=msbm5
\newfam\msyfam
\textfont\msyfam=\tenmsy
\scriptfont\msyfam=\sevenmsy
\scriptscriptfont\msyfam=\fivemsy
\edef\msyfam@{\hexnumber@\msyfam}
\def\Bbb#1{\fam\msyfam\relax#1}
\catcode`\@=\active

\renewcommand{\dj}{\Delta_j}
\newcommand{\dl}{\Delta_l}
\newtheorem{theorem}{Theorem}[section]
\newtheorem{proposition}[theorem]{Proposition}
\newtheorem{lemma}[theorem]{Lemma}
\newtheorem{corollary}[theorem]{Corollary}
\newtheorem{definition}[theorem]{Definition}
\newtheorem{example}[theorem]{Example}
\newtheorem{remark}[theorem]{Remark}

\newcommand{\QE}{\mathscr{Q}_{\rm E}}

\renewcommand{\det}{\delta_{\mu\nu}^\perp}

\newcommand{\eq}[1]{\begin{equation}\label{#1}}
\newcommand{\en}{\end{equation}}
\newcommand{\eqn}{\begin{eqnarray*}}
\newcommand{\enn}{\end{eqnarray*}}
\newcommand{\eqnn}{\begin{eqnarray}}
\newcommand{\ennn}{\end{eqnarray}}
\newcommand{\proof}{{\noindent \it Proof:\ }}
\newcommand{\qed}{\hfill {\bf qed}\par\medskip}
\newcommand{\BR}{{{\Bbb R}^3 }}
\newcommand{\bi}{\begin{description}}
\newcommand{\ei}{\end{description} }

\newcommand{\RR}{{\Bbb R}}
\newcommand{\R}{\RR}

\newcommand{\bl}[1]{\begin{lemma}\label{#1}}
\newcommand{\el}{\end{lemma}}
\newcommand{\bc}[1]{\begin{corollary}\label{#1}}
\newcommand{\ec}{\end{corollary}}
\newcommand{\bt}[1]{\begin{theorem}\label{#1}}
\newcommand{\et}{\end{theorem}}
\newcommand{\bp}[1]{\begin{proposition}\label{#1}}
\newcommand{\ep}{\end{proposition}}
\newcommand{\br}[1]{\begin{remark}\label{#1}}
\newcommand{\er}{\end{remark}}

\newcommand{\limn}{\lim_{n\rightarrow\infty}}

\newcommand{\Je}{{\bf J}}

\newcommand{\W}{{\cal W}}
\newcommand{\kak}[1]{(\ref{#1})}
\newcommand{\LR}{{L^2(\BR)}}

\newcommand{\LRT}{{L^2(\RR^{3 +1})}}

\newcommand{\fff}{\mathscr{F}}
\newcommand{\fffr}{{\fff}} 

\newcommand{\is}{\inf\sigma}

\newcommand{\lk}{\left(}
\newcommand{\rk}{\right)}
\newcommand{\lkk}{\left\{}
\newcommand{\rkk}{\right\}}

\newcommand{\add}{a^{\ast}}

\newcommand{\PF}{H}

\newcommand{\ov}[1]{\overline{#1}}
\newcommand{\hf}{H_{\rm f}}

\newcommand{\gr}{\varphi_{\rm g}}

\newcommand{\grt}{\gr^t}

\newcommand{\half}{\frac{1}{2}}
\newcommand{\han}{{1/2}}
\newcommand{\wick}[1]{:\!\!{#1}\!\!:}

\newcommand{\I}{\mathrm{i}}
\newcommand{\EV}[1]{\E_{\mu_T^V}\left[{#1}\right]}
\renewcommand{\det}{\delta^\perp}

\newcommand{\vvv}[1]{\left[
\!\!\!\begin{array}{c}#1\end{array}\!\!\!\right]}

\newcommand{\CCR}{X}
\newcommand{\la}{\lambda}
\newcommand{\s}{\sigma}
\newcommand{\hhh}{\mathscr{H}}

\newcommand{\db}[1]{\delta B_{#1}}

\newcommand{\jjj}{\sum_{j=1,2}}

\renewcommand{\d}{\displaystyle}
\newcommand{\av}{A} 
\newcommand{\vp}{{\hat \varphi}}

\newcommand{\non}{\nonumber}

\newcommand{\e}{\mathrm{e}}
\newcommand{\D}{\mathrm{d}}

\newcommand{\G}{\mathcal G}
\newcommand{\E}{\mathbb{E}}

\newcommand{\om}{\omega}
  \newcommand {\norm} [2] [] {\ensuremath{ \left\Vert  #2  \right\Vert_{#1} } }
\newcommand{\Z}{\mathbb{Z}}
\newcommand{\eps}{\varepsilon}

\newcommand{\CV}{{\Bbb V}}
\newcommand{\CW}{{\Bbb W}}

\makeatletter \@addtoreset{equation}{section} \makeatother

\title
{\sc Gibbs measures with double stochastic integrals on a path
space}

\author{\small Volker Betz\thanks{Supported by an EPSRC fellowship EP/D07181X/1}
\thanks{Mathematics Institute, University of Warwick,Coventry CV4 7AL, United Kingdom, e-mail:	 v.m.betz@warwick.ac.uk} and
Fumio Hiroshima\thanks{This work is financially supported by
Grant-in-Aid for Science Research (C) 17540181 from JSPS. }
\thanks{
Faculty of Mathematics,
Kyushu University,
Hakozaki, Higashi-ku, 6-10-1,
 Fukuoka 812-8581, Japan, e-mail:hiroshima@math.kyushu-u.ac.jp.}
}
\date{\today}
\begin{document}
\pagestyle{myheadings}
\markboth{Double stochastic integrals}{Double stochastic integrals}

\setlength{\baselineskip}{13.8pt}
\maketitle

\begin{abstract}
We investigate Gibbs measures
relative to Brownian motion in the case when
the interaction energy is given by a double stochastic integral.
In the case when the double stochastic
 integral is originating from the Pauli-Fierz model
 in nonrelativistic quantum electrodynamics,
we prove the existence of its  infinite volume limit.
\end{abstract}

\section{Preliminaries} \label{S1}
\subsection{Gibbs measures relative to Brownian motions}
 Gibbs measures relative to Brownian motion
  appeared
 in \cite{Ne64}, where they have been introduced
to study a particle system linearly coupled to  a scalar quantum field.
 A systematic study of
  such measures has been started from  \cite{bhlms},
   where by making use of this measure the spectrum of
   the so-called  Nelson model is investigated.
 Since then there has been growing activity and interest in the study of various
  types of these measures
  \cite{BeLo03,BS05,Gu06,GuLo07,LoMi}.

One way to understand Gibbs measures relative to Brownian motion is
to view them as the limit of a one-dimensional chain of unbounded
interacting spins, with the distance between the spins going to
zero. As a simple example, which will be instructive in what
follows, let us take $\R^{d}$ for the spin space, and fix a (finite
or infinite) a priori measure $\nu_{0}$ on
$\R^{d}$ as well as
smooth, bounded functions $\CV: \R^{d} \to \R$ and $\CW: \R \times
\R^{d} \to \R$.
 On the lattice $\eps \Z \cap [-T,T]$
 with spacing
$\eps$ and $n = 2T/\eps$ sites,
we define the measure
 $\nu^\CW$
 through
\begin{equation}   \label{dr}
\nu^\CW
 (\D x_{-n} \ldots \D x_{n})
 = \frac{1}{Z_{\nu^\CW}} \prod_{|i| \leq n} \nu_{0}(\D x_{i})
\e^{- \eps \sum_{i} \CV(x_{i})
- \frac{1}{\eps} \sum_{i} (x_{i+1}-x_{i})^{2} +
 \eps^{2} \sum_{i,j} \CW(j-i, x_{j}-x_{i})}.
 \end{equation}
 Here $Z_{\nu^\CW}$ normalizes $\nu^\CW$ to a
probability measure, and for finite $\eps$, $\nu^\CW$ is just a
chain of interacting spins. However, the scaling becomes very
important when $\eps \to 0$. Then, formally each spin
 configuration
$(x_{i})_{|i|\leq n}
\in \eps Z \cap [-T,T]$ becomes a function
$x(\cdot)$ on $[-T,T]$, and  the particular scaling
 of the quadratic
term above gives rise to the term
 $ \lim_{\eps\rightarrow 0}
\eps\sum_i{(x_{i+1}-x_i)^2/ \eps^2}=\int_{-T}^T (\D x(s)/\D
s)^2 \D s$.
 It is this term that prevents the measure
  $\nu^\CW$ from being
concentrated on more and more rough functions when $\eps \to 0$,
ensuring continuity of $x(t)$ in the limit. Indeed, when $\nu_{0}$
is chosen as the Lebesgue measure on $\RR^d$, it is not difficult to show that part
of the normalization along with
 the quadratic term give converge to
Wiener measure $\W$, so that in the limit,
 $\eps\rightarrow 0$,
 we obtain
\begin{equation} \label{double riemann}
\nu^\CW_T(\D B) =
 \frac{1}{Z_{\nu_T^\CW}} \e^{
   - \int_{-T}^{T} \CV(B_{s}) \, \D
s - \int_{-T}^{T}
 \D s\int_{-T}^{T} \D t \CW(B_{t}-B_{s},t-s)}
\, \D \W.
\end{equation}
Here $(B_t)_{t\geq 0}$ is now a Brownian motion, hence we  call
\kak{double riemann} Gibbs measures relative to Brownian motion.
Indeed, the measure appearing in \cite{Ne64} is
 of the above type, and most of the subsequent works cited above have been concerned with measures of the form (\ref{double riemann}).

In this paper we study another type of Gibbs measures, arising from a very similar discrete spin system. Namely,
let us now define
\begin{eqnarray}
&&
\lefteqn {
\nu^{\CW_M}(\D x_{-n} \ldots \D x_{n}) }\non \\
&&
=
 \frac{1}{Z_{\nu^{\CW_M}}} \prod_{|i| \leq n}
 \nu_{0}(\D x_{i})
\exp
\left(- \eps \sum_{i} \CV(x_{i})
 - \frac{1}{\eps} \sum_{i} (x_{i+1}-x_{i})^{2} \right.\non \\
&& \hspace{3cm}\left. +
   \sum_{i,j} (x_{i+1}-x_{i}) \cdot \CW_M(j-i, x_{j}-x_{i}) (x_{j+1}-x_{j}) \right).
\end{eqnarray}
Now $\CW_M$ is a $d \times d$ matrix, but otherwise the expression
looks very similar to \kak{dr}.
 The crucial point is, however, that
now the scaling of the term involving $\CW_M$
 is different. The $\eps^{2}$ which ensured convergence to a double Riemann integral
  is gone by sandwiching $\eps^2 \CW_M$ between
   $(x_{i+1}-x_i)/\eps$ and $(x_{j+1}-x_j)/\eps$,
  and
replaced by the increments of the spins themselves. Since these increments will eventually converge to Brownian motion increments,
as discussed above, they are of order $\sqrt{\eps}$, so the scaling is indeed different. So after taking the limit $\eps \to 0$, we informally obtain
\begin{equation} \label{double stochastic}
\nu^{\CW_M}_{T}( \D B) =
 \frac{1}{Z_{\nu^{\CW_M}_T}} \e^{ - \int_{-T}^{T}
\CV(B_{s}) \, \D s - \int_{-T}^{T} \int_{-T}^{T} \D B_{s} \cdot
 \CW_M(B_{t}-B_{s},t-s) \, \D B_{t}} \D \W.
\end{equation}
As a consequence, taking the limit $\eps \to 0$  yields a double stochastic integral in place of the double Riemann integral \kak{double riemann}.

\subsection{Definition of double stochastic integrals}
 From now on we assume that $d=3$ and specify the pair potential
$\CW_M=W=W(X,t)=(W_{\mu\nu}(X,t))_{1\leq \mu,\nu\leq 3}$ given by
 \eq{w}
W_{\mu\nu}(X,t):=
 \int\frac{|\vp(k)|^2}{2\omega(k)}e^{-\omega(k) |t|}
e^{ik\cdot X} \det_{\mu\nu}(k) \D ^3 k,
 \en
where
  $
    \det(k)=(\det_{\mu\nu}(k))_{1\leq \mu,\nu\leq 3}$ is given by
      \eq{det}
  \det_{\mu\nu}(k)
 := \delta_{\mu\nu} - \frac{k_\mu k_\nu }{|k|^{2}}.
\en
 Measures of the type (\ref{double stochastic}) with pair
potential
 \kak{w}
 appear in the study
of the so-called non-relativistic quantum electrodynamics,  and have
been introduced on a formal level in
 \cite{Fey,ha,sp5}.
 However we notice that
  there are some difficulties  in the expression
  \kak{double stochastic}:
For $t>s$, the integrand is not adapted to the natural filtration
$\fff_T=\s(B_r;r\leq T)$, so as a stochastic integral or any of its
obvious transformations the double stochastic integral such as
 \kak{double stochastic} does not make sense.
 So the right-hand side of \kak{double stochastic} is just
 an informal symbol.

 In \cite[Definition 4.1]{h13} and \cite[(4.2)]{h9}, however,  the firm mathematical
 definition of \kak{double stochastic} has been given
 through a Gaussian random process associated with
 an Euclidean quantum field. We outline it below.
 A Gaussian random
 process $\mathscr{A}^E(f)$ labeled by $f\in \oplus^3 \LRT$
 on some probability space
   $(\QE,\Sigma_E,\mu_E)$
 is introduced, which has mean zero and covariance
$\E_{\mu_E}[\mathscr{A}(f)\mathscr{A}(g)]=q(f,g)$ given by
 \eq{q}
 q(f,g):=\half \int\ov{\hat f(k,k_0)}
 \cdot \det(k) \hat g(k,k_0)\D ^3 k \D k_0
 \en
 for $f,g\in \oplus^3 \LRT$, where $\hat {\ }$ denotes the Fourier transformation.
 Let
  \eq{st}
   K_t=\oplus_{\mu=1}^3 \int_0^t j_s\varphi(\cdot-B_s)
   \D B^\mu_s\en
    be  the $\oplus^3 \LRT$-valued stochastic integral defined in the
similar way as
  standard stochastic integrals,
  where
  $j_s:\LR\rightarrow \LRT$ denotes the isometry satisfying
   \eq{js}
   (j_sf, j_t g)_\LRT =(\hat f, e^{-|t-s|\omega}\hat g)_\LR.
   \en
   See Subsection 3.1 for the details.
 \begin{definition}\label{def}
 Let $W$ be the pair potential defined in \kak{w}.
 The dounle stochastic integral 
is defined by
   \eq{sta}
  \int_0^t \int_0^t dB_s\cdot W(B_t-B_s,t-s)dB_t:=q(K_t,K_t).
   \en
\end{definition}
 We would like to express \kak{sta} as an iterated
 stochastic integral in this paper.

 \subsection{Main results}
 Let us define the Wiener measure $\W$ on $\CCR:=C(\RR;\BR)$, cf.\ also \cite[p. 39, Remark 1.]{Si82}.
Let
$H_{0}=-(\han)\Delta$.
 Suppose that
 $f_1,...,f_{n-1}
 \in L^\infty(\BR)$ with compact support.
 Then
  there exists a measure
   $\W$ on $\CCR $ such that
 \begin{eqnarray}
   \lefteqn{ (f_{0}, e^{-(t_1-t_0) H_0}f_1
   \cdots
 f_{n-1} e^{-(t_n-t_{n-1})H_0}
  f_n)_{\LR} } \nonumber \\
    &=& \int _{\CCR }
    \ov{f_0(B_{t_0})} f_1(B_{t_1})
  \cdots f_n(B_{t_n})
   \D \W. \label{win}
 \end{eqnarray}
A path with respect to this measure is
denoted by $B_{t}(w)=w(t)$ for
$w\in \CCR $. Note that Wiener measure is not a probability measure, indeed it
 has infinite mass.
If $P_W^{x,t_{0}}$ denotes the measure of standard Brownian motion starting from $x\in\BR$ at time $t_0$, then
\[
   \int _{\CCR }
   f_0(B_{t_0})
  \cdots f_n(B_{t_n})
   \D \W
   = \int_\BR dx \int _{C([t_0,\infty);\BR)} f_0(B_{t_0})
  \cdots f_n(B_{t_n})
     \D P_W^{x,t_{0}} .
\]Let $\psi\in\LR$ be a nonnegative function and we fix it throughout this paper.
 In the case of \kak{double riemann}, the
existence of the weak limit of the  measure on $\CCR$,
$$d\nu^\CW_T:=
\frac{1}{Z_{\nu_T^\CW}}\psi(B_{-T})\psi(B_T) e^{-\int_{-T}^T ds\int_{-T}^T dt
 \CW(B_t-B_s,t-s)}
  e^{-\int _{-T}^T \CV(B_s) ds}
 \D \W, $$ as
$T\rightarrow \infty$ has been investigated
 for various kinds of
$\CV$ and $\CW$, and the limiting measure, $\nu^\CW_\infty$,
 proved to be useful to study
 the ground state $\gr$ of some particle system linearly coupled to
 a scalar quantum field.
  Namely for a suitable operator ${\cal O}$,
 we can express the expectation $(\gr, {\cal O}\gr)$ as
  $\int_\CCR f_{\cal
  O}d \nu^\CW_\infty$ with some integrand $f_{\cal O}$.
 So, beyond the existence of a measure of the form
 (\ref{double stochastic}), one is interested in the limit as $T \to \infty$, at
least along a subsequence.
 In other words, one would like to prove the tightness of
 the family of measures \kak{double stochastic}.
       This is by no means an easy task, given that there are very few good general estimates on single stochastic integrals, let alone double integrals.

 The purpose of our present paper is to point out that there
 is at least one special case where there is a comparatively easy way
 to construct both the finite volume Gibbs measure
 and the infinite volume limit,
 namely the case when
  $\CW_M=W=(W_{\mu\nu})_{1\leq\mu,\nu\leq 3}$
    is given in \kak{w}.
     Fortunately,
 this special case is the one that motivated
 the whole theory of Gibbs measures with double stochastic integrals.
 The main results in this paper are\\
(1) we give an iterated stochastic integral expression of
 \kak{sta};\\
(2) we show the tightness of the family of measures
\[
  \frac{1}{Z_T}
 \psi(B_T)\psi(B_{-T}) \e^{
  -\int_{-T}^T V(B_s) ds-\alpha^2 \int_{-T}^T \int_{-T}^T dB_s\cdot W(B_t-B_s,t-s)\D B_t}
   \D \W
 \]
  for a general class of $V$ including the Coulomb potential $V(x)=-1/|x|$, and arbitrary values of coupling constant $\alpha\in\RR$.

There has been recent progress both of the above topics:
 M. Gubinelli and J. L\H{o}rinczi \cite{GuLo07} employ
 the concepts of stochastic
currents rough paths in order to define (\ref{double stochastic})
rigorously for finite volume, and use a cluster expansion in order
to construct the infinite volume limit. While these are impressive
results, the techniques used are rather advanced, and
 the use of cluster expansion comes with
  strong assumptions
 on single site potentials  $V$ and  coupling constants.
The advantage of our methods is
    that we can avoid some restrictions
     needed in \cite{GuLo07}; in particular
     we need not restrict to single site potentials
     that grow faster than quadratically at infinity,
     and we need no small coupling constant
     in front of the double stochastic integral.
      In particular our results include the Coulomb potential
      which is the most reasonable  single site potential.
      On the other hand, of course the range of potentials
       $W$ that is treated in \cite{GuLo07} is much greater than ours.

       The paper is organized as follows:
In Section 2 we will construct the finite volume Gibbs measure
 as the marginal of a measure with single stochastic integral
 on a larger state space.
 This construction is well known \cite{Sp04},
  but has not been carried out rigorously so far.
  In Section 3, we rely on the detailed results available
  about the Pauli-Fierz model
  \cite{h26,gll} in order
  to show that our family of Gibbs measures is tight,
  giving the existence of an infinite volume measure.
  While we expect that the general method
  of enlarging the state space should allow us to define and prove infinite
  volume limits for many more models than just Pauli-Fierz,
  this is not all straightforward.
  We will comment on this issue at the end of Section \ref{ivl}.

\section{Iterated expression of finite volume measures}

In this section we will specify the measure $\mu_{T}$ that we are working with, and identify it as the marginal of another measure
$\nu_{T}$ on a larger state space.
Let us start by introducing an infinite dimensional
Ornstein-Uhlenbeck process which will serve as the reference measure
for the auxiliary degrees of freedom. Put
 \eq{dispersion}
 \omega(k) =
\sqrt{|k|^{2}+m^{2}} \en
  for $m \geq 0$,
 and let $X_s(f)$
be the Gaussian random process on a probability space
$(\mathscr{Q},\Sigma,{\cal G})$
 labeled by measurable function $f=(f_1,f_2,f_3)$
  with mean zero
  and covariance given by
\eq{cov}
  \E_{\G}[X_{s}(f) X_{t}(g)] = \int \D^{3} k
\frac{1}{2\omega(k)} \e^{-\omega(k)|t-s|}
 \ov{\hat{f}(k)} \cdot
 \det(k) \hat{g}(k).
\en
  Here $\hat f$ denotes the Fourier transform of $f$ and
  we assume that $\hat f_\mu/\sqrt\omega, \hat g_\nu/\sqrt\omega\in\LR$, $\mu,\nu=1,2,3$.
  \begin{remark}{\rm
  \label{R}
Let $Y_s(f)$ be the Gaussian random process on
$(\QE,\Sigma_E,\mu_E)$ defined by
 \eq{ys}
 Y_s(f):=\mathscr{A}^E(j_s
(\hat f /\sqrt\omega)^\vee).
\en
 Then $Y_s(f)$  is mean zero and
its covariance is
 \eq{YY}
  \E_{\mu_E}[Y_s(f)Y_t(f)]=\E_\G[X_s(f)X_t(g)].
  \en
   Hence $Y_s(f)$ and $X_s(f)$ are isomorphic as Gaussian random processes.
 } \end{remark}

 We will now couple $\G$ to the Wiener measure $\W$.
 For this we use a coupling function
 $\varphi$ with the assumption below:\\[5mm]
{\bf Assumption (A):}
 \begin{itemize}
 \item[(1)] $\vp(k) = \vp(-k) = \overline{\vp(k)}$
 and
 $ \sqrt{\omega}
 \vp, \vp/\omega \in L^{2}(\R^{3})$.
 \item[(2)]
  $\vp$ is rotation invariant, i.e.,  $\vp(Rk)=\vp(k)$ for all $R\in
 O(3)$.
\end{itemize}

Let us now define the quantity
\eq{defj}
J_{[0,T]}(X):=\int_{0}^{T}
  X_{s}(\varphi(\cdot-B_{s})) \cdot \D B_s.
  \en
  The proper definition of $J_{[0,T]}$ reads
\begin{equation} \label{stoch int def}
J_{[0,T]}(X)
 := \lim_{n \to \infty}
 \sum_{j=1}^{n}  X_{(j-1)T /n}(\varphi(\cdot -B_{(j-1)T/n})  (B_{jT /n}-B_{(j-1)T/n})),
\end{equation}
where the right hand side strongly converges in
 $L^{2}(\CCR \times \mathscr{Q} ; \G \otimes
P_{W}^{x,0})$.
This is proved by showing that the right-hand side of \kak{stoch int def} 
is Cauchy by making use of \kak{cov}. 
In the same way, we can define
\eq{defjj}
J_{[-T,T]}(X):=\int_{-T}^{T}
  X_{s}(\varphi(\cdot-B_{s})) \cdot \D B_s,
  \en
  where $(B_t)_{t\in\RR}$ denotes the $3$-dimensional
   Brownian motion on the whole time line $\RR$. The coupling between the Gaussian process and Brownian motion is  given by the measure $\nu$ on $\CCR \times
\mathscr{Q} $ with
\begin{equation}
 \label{nut}
\D \nu_{T} = \frac{1}{Z_{T}}
  \exp \left(
  \I \alpha \int_{-T}^{T}   X_{s}(\varphi(\cdot-B_{s})) \cdot \D B_s \right) \psi(B_{-T}) \psi(B_{T})
   \D \W\otimes \D \G,
\end{equation}
where $\psi\in\LR$ is an arbitrary nonnegative function, $Z_T$ the normalizing constant, and $\alpha$ is a coupling constant.
 In order to guarantee that the density in (\ref{nut}) is
integrable with respect to $\W$, we chose the boundary function
$\psi$ to be of rapid decrease at infinity.

We are now in the position to define our finite volume Gibbs measure.
We will introduce an
on-site potential $V$ which we take Kato-decomposable \cite{BHL00},
i.e.\ we require that the negative part $V_{-}$ is in the Kato class
while the positive part $V_{+}$ is the locally Kato class
\cite{Si82}. This ensures e.g.\
 that
 \eq{kato}
 \sup_{x} \E_{P_{W}^{x,0}}\left[ \exp\lk -\int_{0}^{t}
V(B_{s}) \, \D s\rk\right]<\infty.
\en

\begin{definition}
 \label{dd}
 Let $V: \R^{3} \to \R$ be Kato-decomposable and $\alpha\in\RR$ a coupling constant.
 Then
 the measure $\mu_{T}^{V}$ on $\CCR$
 is defined through
 \eqnn
&&
  \D \mu_T^{V}:= \frac{1}{Z_{T}}
  \e^{-\int_{-T}^T  V(B_{s}) \, \D s} \E_{\cal G}[\D \nu_T]\non\\
  &&
  =
    \frac{1}{Z_{T}}
  \psi(B_{-T}) \psi(B_{T})
  \e^{-\int_{-T}^T  V(B_{s}) \, \D s}
 \E_\G\left[
   \exp \left( \I \alpha
\int_{-T}^{T}
  X_{s}(\varphi(\cdot -B_{s}))
\cdot \D B_{s}\rk\right]
 \D
\W .\non\\
&&\label{meas} \ennn
\end{definition}

  We want to show that the measure $\mu_{T}^{V}$ we just defined is
  a Gibbs measure with double stochastic integral as given in Section \ref{S1}. The key to doing this is the fact
  that we will be actually able to calculate the Gaussian integral
  $\int_\mathscr{Q}   \exp(\I  J_{[-T,T]}(X)) \D \G(X)$,
  and thus are left with an expression involving Brownian motion paths
only. In doing so, we will set $\alpha=1$ for a simpler notation.

Let us give the heuristic presentation first.
 By the standard formula we have
\begin{equation} \label{formal int}
\E_{\G}[\e^{\I J_{[-T,T]}}] =
 \exp \left( -\frac{1}{2} \E_{\G}[J_{[-T,T]}^{2}]
\right) \en
 and formally, by Remark \ref{R}, we have
 \eq{formal int2}
 \E_{\G}[J_{[-T,T]}^{2}]\stackrel{\rm formal}{=}
 \frac{1}{2} \int_{-T}^{T}
\int_{-T}^{T} \D B_{s} \cdot W(B_{t}-B_{s},t-s) \D B_{t},
  \en
where $W$ is given in \kak{w}.
%
 As it stands, there are problems with the right-hand side of
 formal expression  (\ref{formal int2}),
 mainly because the integrand is not adapted.
 The resolution is to use symmetry of $W$ and
break up the integral into two parts, one where $s<t$ and one where
$s>t$, which are then proper iterated It\^o integrals.
 This leaves the diagonal part,
which gives a non-vanishing contribution
 by the unbounded variation
of  $B_{t}$.

We define the iterated stochastic integral $S_T$ by
\begin{eqnarray}
S_T &:=&
    \int \D^3 k \frac{|\vp(k)|^{2}}{2\om(k)}
   \int_{-T}^{T} \e^{\I k\cdot B_{s}} \D B_{s} \cdot
   \int_{-T}^{s}
   e^{-\om(k)(s-r)} \e^{-\I k\cdot  B_{r}}\det(k) \D B_{r}
   + \nonumber \\
   && + \frac{T}{3} \int \D^3 k \frac{|\vp(k)|^{2}}{2 \om(k)}
   \label{25}
\end{eqnarray}
$S_{T}$ is the well-defined expression that will replace
(\ref{meas}). The above line of reasoning and (\ref{25}) are not new \cite{Sp04}, except that (\ref{25}) is usually not written out but instead just referred to as the double stochastic integral with the diagonal removed. Nevertheless, (\ref{25}) can be considered as known. However, the derivation above is mathematically not rigorous, since the ill-defined expression (\ref{formal int2}) appears along the way. To avoid this, one has to derive (\ref{25}) directly from
$\E_{\G}[\e^{\I J_{[-T,T]}}]$. This is what we do in the next theorem.

\bt{integrate out}
 For almost every $w\in\CCR$,
 we have
\begin{eqnarray}
 \label{ds}  \E_{\G} [ \e^{\I J_{[-T,T]}}]= e^{-S_{T}}.
\end{eqnarray}
\et
 \proof
 Let us replace the time interval $[-T,T]$
  with $[0,T]$ for notational convenience.
 We employ (\ref{stoch int def})
  and use dominated convergence to get
\begin{eqnarray*}
\E_{\G}[e^{\I J_{[0,T]}}]
 &=& \lim_{n \to \infty} \E_{\G}
 \left[ \exp \left( {\I \sum_{j=1}^{n} X_{\dj }
 (\varphi(\cdot-B_{\dj })) \cdot \db{j} }\right)\right]\\
&=& \lim_{n \to \infty}
 \exp \left( - \frac{1}{2} \E_{\G}
 \left[ \sum_{j=1}^{n } X_{\dj } (\varphi(\cdot-B_{\dj }))
 \cdot \db j  \right]^{2} \right),
\end{eqnarray*}
where we set
 $\db j= B_{jT/n}-B_{(j-1)T/n}$ and $\dj=(j-1)T/n$, $j=1,...,N$.
 Now
\eqnn
 \lefteqn{\E_{\G} \left[ \sum_{j=1}^{n}  X_{\dj }
(\varphi(\cdot-B_{\dj })) \cdot \db j  \right]^{2} } \non
\\
&=& \int \D^3 k \, \frac{|\vp(k)|^{2}}{2 \om(k)} \sum_{j=1}^n
\sum_{l=1}^n  \e^{-{|\dj-\dl|\om(k)}}
\e^{\I k (B_{\dj}- B_{\dl})}
\db j\cdot
 \det(k)  \db l  \non
 \\
&=&
 \label{mo2} 2 \sum_{j=1}^{n}
 \int \D^3 k \,
 \frac{|\vp(k)|^{2}}{2
\om(k)}
 e^{-\dj\omega(k)} e^{\I k B_{\dj}} \sum_{l=1}^{j-1} \e^{+\dl \om(k)}
\e^{-\I k B_{\dl }} \db j  \cdot \det(k) \db l   \\
&&
\label {mo} \hspace{4cm}
 + \sum_{j=1}^{n}
 \db j
 \cdot \lk
 \int \D^3 k \,
\frac{|\vp(k)|^{2}}{2 \om(k)}
 \det(k) \rk \db j .
\ennn
For the diagonal term in the last line above
 we   note that
 $$\int
\frac{|\vp(k)|^{2}}{2 \om(k)} \det_{\mu\nu}(k)\D^3 k
 = \delta_{\mu\nu}\frac{2}{3}
\int  \frac{|\vp(k)|^{2}}{2 \om(k)}\D^3 k $$ by the rotation
invariance of $\vp$.
Now as
 $n \to \infty$,
 $\sum_{j=1}^{n} |\db j|^{2} \to T$,
while $\sum_{j=1}^{n} |\db j ^{\mu}|^{2} \to T/3$ for all
$\mu=1,2,3$,
for almost every $w\in\CCR$.
 Thus
 for almost every $w\in\CCR$, we find
\[
\lim_{n \to \infty} \sum_{j=1}^{n} \db j
 \cdot \lk
 \int \D^3 k \,
\frac{|\vp(k)|^{2}}{2 \om(k)}
 \det(k)\rk
 \db j  = \frac{2T}{3} \int \D^3 k \,
\frac{|\vp(k)|^{2}}{2 \om(k)}.
\]
For the off-diagonal term, we start by noting that by the definition
of the It\^o integral for  locally bounded functions $f,g: \R \times
 \R^{3} \to \R$,
  we can see that
$$\E_{P_W^{0,0}}\left
[\int_0^t ds \left| f(s,B_s)\int_0^s g(r,B_r)\D B_r^\mu \right|^2\right]
<\infty.$$
 Hence the stochastic integral of
 $\rho^\mu(s)=
 f(s,B_s)\int_0^s g(r,B_r)\D B_r^\mu$ exists for all $\mu=1,2,3$, and it holds that
for each $k\in\RR^3$,
\begin{eqnarray}
&&
  \limn \sum_{j=1}^{n}
 \lk
 f(\dj ,
B_{\dj })
\db j \cdot \delta^\perp(k)  \int_{0}^{\dj }
g(r,B_{r})  \D B_{r} \rk \non \\
&&\label{double 1}
 =
 \int_{0}^{T}
  f(s,B_{s})\D B_s\cdot\delta^\perp(k) \lk
 \int_{0}^{s} g(r,B_{r}) \, \D B_{r}\rk
 \end{eqnarray}
  strongly in $L^2(\CCR;P_W^{0,0})$.
 By the independence of
 Brownian increments and the fact that
  $\E_{P_W^{0,0}} [(\db j )^2] = 1/n$, $\E_{P_W^{0,0}}[\db j ]=0$,
   we can estimate the
 $L^{2}(\CCR;P_W^{0,0})$-difference of (\ref{double 1})
 and the off-diagonal term:
\begin{eqnarray}
\lefteqn{\E_{P_W^{0,0}}
\left[ \sum_{j=1}^{n}
 f(\dj ,B_{\dj })
 \db j\cdot \delta^\perp(k) \left( \int_{0}^{\dj } g(r,B_{r}) \, \D B_{r} -
 \sum_{l=1}^j  g(\dl , B_{\dl }) \db l  \right)
   \right]^{2} } \non \\
&=& \frac{2}{3}
\frac{1}{n} \sum_{\nu=1}^3
\delta_{\nu\nu}(k) \E_{P_W^{0,0}}
 \left[
  \sum_{j=1}^{n} f(\dj ,B_{\dj })^{2}
   \left(
    \int_{0}^{\dj }
     g(r,B_{r}) \, \D B_{r}^\nu -
      \sum_{l=1}^j  g(\dl , B_{\dl })
      \db l^\nu  \right)^{2}
  \right] \non\\
&\leq&
\label{non} \norm[\infty]{f}^{2}
\frac{2}{3} \frac{1}{n}
 \sum_{\nu=1}^3
\delta_{\nu\nu}(k)
\sum_{j=1}^{n}
\E_{P_W^{0,0}}
 \left[
 \int_{0}^{\dj } g(r,B_{r}) \, \D B_{r}^\nu -
\sum_{l=1}^j  g(\dl , B_{\dl })  \db l^\nu  \right]^{2}.
\end{eqnarray}
Then the right-hand side above converges to zero as
$n\rightarrow\infty$,
and together with \kak{double 1} we find
\begin{eqnarray}
&&
\limn \sum_{j=1}^n f(\dj, B_{\dj})\db j\cdot \delta^\perp(k)\sum_{l=1}^j g(\dl, B_{\dl})\db l \non \\
&&\label{r1}
=
\int f(s,B_s)\D B_s\cdot \delta^\perp(k)\int_0^s g(r,B_r) \D B_r
\end{eqnarray}
 in $L^2(\CCR;P_W^{0,0})$.
 By putting $f(t,x) = \e^{\I k\cdot  x}
  \e^{-\om(k) t}$ and  $g(t,x) =
\e^{- \I k \cdot x} \e^{\om(k) t}$ in \kak{r1},
we can see that \kak{mo2} converges to the off-diagonal part of $S_T$.
Then  the   proof is finished. \qed
 \begin{remark}\label{unit}
 {\rm It is interesting that we know that $|e^{-S_{T}}| = |\E_{\cal G}[e^{iJ_{[-T,T]}}]|
\leq 1$ almost surely.
 This is not obvious from the iterated integral representation
 $e^{-S_T}$.
}\end{remark}
Let us summarize:
\begin{proposition} \label{mut}
Let $\mu_{T}^{V}$ be the measure on $\CCR$ from Definition \ref{dd}. Then
 \[
\D \mu^V_T = \frac{1}{Z_T}
 \psi(B_{-T})\psi(B_T) \e^{-\alpha^2 \hat S_{T}}
\e^{-\int_{-T}^T  V(B_{s}) \, \D s}  \,
 \D \W,
 \]
where
$\hat S_T$ is defined by
$S_T$ with the diagonal part removed:
$$\hat S_T:=
    \int \D^3 k \frac{|\vp(k)|^{2}}{2\om(k)}
   \int_{-T}^{T} \e^{\I k\cdot B_{s}} \D B_{s} \cdot
   \int_{-T}^{s}
   e^{-\om(k)(s-r)} \e^{-\I k\cdot  B_{r}}\det(k) \D B_{r}.$$
Or
$$\hat S_T:=
\int_{-T}^ T Z(s,w)\cdot \D B_s, $$
where
$$Z(s,w)=\int_{-T}^s \D B_r
  \lk
\int\frac{|\vp(k)|^2}{2\omega(k)}\det(k)e^{-(s-r)\omega(k)}e^{-ik\cdot(B_r-B_s)}\D^3 k\rk .$$
 \end{proposition}
\begin{remark}
{\rm
In Proposition \ref{mut},
the diagonal term $
\frac{T}{3} \int \D^3 k \frac{|\vp(k)|^{2}}{2 \om(k)}$ is
absorbed in the normalization constant, since it does not depend on
the Brownian path $B$. Moreover from  Remark \ref{unit} it follows that
$$\left|\exp\lk {-\hat S_T}\rk \right|\leq \exp\lk {\frac{T}{3} \int \D^3 k \frac{|\vp(k)|^{2}}{2 \om(k)}}\rk.$$
Thus $\hat S_T$ may be written symbolically as
$$\hat S_T=\half \int_{[-T,T]\times[-T,T]\setminus\{s=t\}}
\D B_s\cdot W(B_t-B_s,t-s)\D B_t.$$
}\end{remark}

\section{The infinite volume limit} \label{ivl}
\subsection{Tightness and the Pauli-Fierz model}
The idea of the proof of the infinite volume limit we are about to
give is  not straightforward.
 We will show that it follows from showing  that the bottom of
 the spectrum
 of a self-adjoint operator is eigenvalue.
 Actually, in the case of pair potential $W$ under consideration,
   associated  self-adjoint operator is realized
   as the Pauli-Fierz Hamiltonian $H$ in
   the non-relativistic quantum electrodynamics.
   Fortunately it is established that $H$ has the unique ground state
   for not only confining external potential $V$, e.g., $V(x)=|x|^2$,
    but also
   the Coulomb  $V(x)=-1/|x|$,
   which is the most important case.

Let us begin with defining the Pauli-Fierz Hamiltonian
  with  form factor $\vp$
 as a  self-adjoint operator on some Hilbert space $\hhh$ and we
 will review  the functional integral representation of
  the $C_0$ semigroup $e^{-tH}$.

Let
$\fffr:=\bigoplus_{n=0}^\infty [\bigotimes_s^n
 L^2(\BR\times\{1,2\})]$ be the Boson Fock space.
The state space of one electron minimally coupled with the photon (bose) field is
given by
\eq{defhi}
\hhh:=\LR\otimes \fffr.
\en
  We denote the formal kernels of the annihilation operator and the
creation operator on $\fffr$ by $a(k,j)$ and $\add(k,j)$,
respectively, which satisfy the canonical commutation relations:
\eq{abc}
[a(k,j),\add(k',j')]=\delta(k-k')\delta_{jj'},\quad [a(k,j),a(k',j')]=0=[\add(k,j),\add(k',j')].
\en
 The free
Hamiltonian in $\fffr$ is defined by
\eq{defhf}
\hf:=\jjj\int \omega(k) \add(k,j) a(k,j) \D^3 k.
\en
Here dispersion relation $\omega$ is given by \kak{dispersion}.
 Let us fix a function $\vp$ satisfying Assumption (A)
The
 quantized radiation field $\av=(\av_1,\av_2,\av_3)$ with form factor $\vp$
 is
 defined by
 $
 \av_\mu :=\int_\BR^\oplus \av_\mu (x)
\D^3 x$,
where we used  the isomorphism $\hhh\cong \int_\BR^\oplus \LR\  \D x$ and
\eq{defav}
\av_\mu (x)
:=\frac{1}{\sqrt2}\sum_{j=1,2} \int e_\mu(k,j)\lk e^{-ikx}
\frac{\vp(k)}{\sqrt\omega(k)}\add(k,j)
 +e^{ikx}\frac{\vp(-k)}{\sqrt{\omega(k)}}a(k,j)\rk \D^3 k.
 \en
 The vectors  $e(k,j)$, $j=1,2$, are the polarization vectors.
 They satisfy $e(k,i)\cdot e(k,j)=\delta_{ij}$ and
 $k\cdot e(k,j)=0$.
 Note that  \eq{pol}
 \jjj  e_\mu (k,j) e_\nu(k, j)=
 \det _{\mu\nu}(k).
 \en
 \kak{pol} is of course independent of the choice of
 polarization vectors and
 $k\cdot e(k,j)=0$ yields that
 \eq{coulomb}
\sum_{\mu=1}^ 3 \nabla_{x_\mu} A_\mu(x)=0.
\en
  The  Pauli-Fierz Hamiltonian $\PF(0)$ is defined by
 \eq{pf}
 \PF(0):=\half(-i\nabla \otimes 1 -\alpha \av )^2+1\otimes
 \hf,
 \en
where $\alpha\in\RR$ denotes coupling constant.
 It is established in \cite{h11,h16}
 that
 $\PF(0)$ is self-adjoint on
$D(-\Delta)\cap D(\hf)$ and bounded from below. Moreover $\PF(0)$ is essentially self-adjoint on any core of $-(\han)\Delta\otimes 1+1\otimes \hf$.
We now introduce a class of external potentials $V:\BR\rightarrow \R$
that we can add to $H_{0}$.
 \begin{definition}
 \label {v}
  $V\in K$ if and only if $V=V_+-V_-$ such that $V_\pm\geq 0$, $V_+\in L_{\rm loc}^1(\BR)$ and $V_-$ relatively form bounded with respect to $-(\han)\Delta$ with bound strictly smaller than one.
   \end{definition}
  Let $V\in K$. Then we define $\PF$ as
    \eq{hv} \PF:=
    \PF(0)\,\, \dot+\,\,  V_+\otimes 1
    \,\,\dot -\,\, V_-\otimes 1,
    \en
where $\dot\pm$ denotes the quadratic form sum.
To see the weak convergence of $\mu_T^V$, we introduce the assumption below.\\[5mm]
{\bf Assumption (GS):}
 There exists a ground state $\gr$ of $H$.
\begin{example}
{\rm
Let
    \eq{cv}
    V(x)=-\frac{C}{|x|} +U(x),
    \en
     where $C\geq 0$ is a constant,  and $U=U_+-U_-\in L_{\rm loc}(\BR)$ such that $U_\pm\geq 0$,
      $\inf _{x\in\BR}U(x)>-\infty$,
        $U_-$ is compactly supported, and $-(\han)\Delta +U$ has a ground state $\phi>0$ with ground state energy $-e_0<0$ such that
         $|\phi(x)|\leq\gamma e^{-|x|/\gamma}$ with some constant $\gamma>0$.
    Then
        the ground state of $H$ exists for
 arbitrary values of  $\alpha$.
 See \cite[p.8]{af} and \cite{bfs3,gll}. Typical examples are
 \begin{eqnarray*}
 && V_{\rm Coulomb} (x)=-\frac{C}{|x|} ,\\
 && V_{\rm confining}(x)=|x|^{2n}, \quad n=1,2,...
 \end{eqnarray*}
}       \end{example}
             To construct the functional integral representation of $e^{-t\PF}$
  we introduce some probabilistic notation which was already mentioned in Section 1.
  Let $\lkk \mathscr{A}^E(f)\rkk_{f\in \oplus^3 \LRT}$, denote the Gaussian random process labeled by $f\in \oplus^3 \LRT$  on some probability space
 $(\QE,\Sigma_E,\mu_E)$ with mean zero and covariance given by
 $\E_{\mu_E}[\mathscr{A}^E(f)\mathscr{A}^E(g)]=q(f,g)$,
  where $q(\cdot,\cdot)$ is defined in \kak{q}.
  We define the isometry
 $j_s:\LR\rightarrow \LRT$ by
$  \widehat
{j_s f}(k,k_0):=
({e^{-ik_0
t}}/{\sqrt\pi})\sqrt{
 {\omega(k)}/({\omega(k)^2+|k_0|^2})} \hat f(k)
$ which satisfies \kak{js}.
 The crucial identity linking
 the Pauli-Fierz model to
 Gibbs measures is
\begin{proposition} \label{funIntRep}\ \\
(1) For arbitrary $f\in\LR$ with $f\geq0$ but $f\not\equiv 0$,
 it follows that
  \eq{pos}
 (\gr, f\otimes\Omega)_\hhh >0
 \en
(2)
Let $f_1,...,f_{n-1}\in L^\infty(\BR)$ and $\psi\in\LR$.
For
$-T=t_0\leq t_1\leq \cdots\leq t_n=T$,
 the Euclidean $n$-point green
function is expressed as
\[
 \frac{
 \lk
 \psi\otimes \Omega,
 e^{-(t_1-t_0)\PF}
 (f_1\otimes1)
 \cdots (f_{n-1}\otimes1)
 e^{-(t_n-t_{n-1})\PF} \psi\otimes \Omega\rk_\hhh}{
  (\psi\otimes\Omega,
 e^{-2TH} \psi\otimes \Omega)_\hhh}
 =\E_{\mu^V_T}
 \left[
    \prod_{j=1}^{n-1} f_j(B_{t_j})\right]
\]

\end{proposition}
\proof
 See \cite{h9} for (1).
 In \cite{h4,h26}
 it is established that
\eqn &&
  (\psi\otimes \Omega,
 e^{-(t_1-t_0)\PF}
 (f_1\otimes1)
 \cdots (f_{n-1}\otimes1)
 e^{-(t_n-t_{n-1})\PF}
 \psi \otimes \Omega)_\hhh\\
 &&
=\E_\W\left[
       \psi (B_{-T}) \psi (B_{T})
    \lk
    \prod_{j=1}^{n-1} f_j(B_{t_j})\rk
e^{-\int_{-T}^TV(B_s)ds} \E_{\mu_E}\left[
 e^{
 -\I \alpha \int_{-T}^{T}
   \mathscr{A}_s^E \cdot \D B_s}\right]\right]
   ,
 \enn
  where
    $$
 \mathscr{A}_{s,\mu}^E
 :=\mathscr{A}^E \lk\frac{}{}\!\!
 \oplus_{\nu=1}^3
 \delta_{\nu\mu} j_s \la(\cdot-B_s)\rk ,\quad \mu=1,2,3,\quad s\in\RR,$$ and $
 \la=(\vp/\sqrt\omega)^\vee.
 $
Since
  $Z_T=(\psi\otimes\Omega,
 e^{-2TH} \psi \otimes \Omega)_\hhh$
  and
 $\d e^{-\alpha^2 S_T}=\E_\G[e^{\I \alpha J_{[0,T]}}]=
\E_{\mu_T}\left[  e^{
 -\I \alpha \int_{-T}^{T}
   \mathscr{A}_s^E \cdot \D B_s}\right]$,
        the lemma follows.
 \qed
 \begin{remark}
 {\rm
 Formally, (2) of Proposition \ref{funIntRep}
 can be deduced from using the Feynman-Kac-It\^o formula \cite{ffg,ha,h4,h26,Si82,Sp04} but note that
 integrand $\mathscr{A}_s^E$ depends on time $s$ explicitly;
although  this formula would give the Stratonovitch integral
$\int_S^T \mathscr{A}_s^E\circ \D B_s=\int_S^T \mathscr{A}_s^E\cdot \D B_s-\half
\int_S^T \nabla \cdot \mathscr{A}_s^E \D s$ instead of the It\^o integral
$\int_S^T
\mathscr{A}_s^E\cdot \D B_s$ above, the Coulomb gauge \kak{coulomb}
 allows us to use the It\^o integral instead, since
  $\nabla_x \cdot \mathscr{A}_s^E(\la(\cdot-x))=0$.
}\end{remark}
 By \kak{pos}, we know that the the ground state, $\gr$,  of $\PF$ is unique if it exists and, in particular,  $(\gr, f\otimes\Omega)_\hhh\not=0$ holds,
  then we can
define the sequence converging to the normalized ground state $\gr$
by
$$\grt:=
  \|e^{-t\PF}(f\otimes \Omega)\|_\hhh^{-1}
  {e^{-t\PF}(f\otimes \Omega)}.
 $$ Actually,
by virtue of \kak{pos},
 we see that
 \eq{2} \gr=s-\lim_{t\rightarrow
\infty} \gr^t. \en
 One immediate and useful corollary
 of \kak{2} and Proposition \ref{funIntRep} is as follows.
  \bc{exp} Let $\rho, \rho_1,\rho_2\in L^\infty(\BR)$.
  Then for $t>s$,
  \eqn
  &&
  \lim_{T\rightarrow\infty}\E_{\mu_T^V}[\rho(B_0)]=(\gr,
  (\rho\otimes 1)\gr)_\hhh,\\
 &&
 \lim_{T\rightarrow\infty}\E_{\mu_T^V}[\rho_1(B_s) \rho_2(B_t)]
  =(\gr,
  (\rho_1\otimes 1)e^{-(t-s)H}(\rho_2\otimes 1) \gr)_\hhh e^{(t-s)E(H)},
\enn where $E(H)=\is (H)$ denotes the ground stare energy of $H$. \ec
In order to prove the main theorem,
 we show  a more general formula than (2) of Proposition \ref{funIntRep}.
 Let
 $$A(\hat f)=
 \frac{1}{\sqrt 2}
 \sum_{\mu=1}^ 3
 \jjj
 \int e_\mu(k,j)
 \lk
 \hat f_\mu (k) \add(k,j)+\hat f_\mu(-k) a(k,j) \rk dk.$$
 Define the family of isometries 
 $\Je_t:\fff\rightarrow L^2(\QE)$, $t\in\RR$,
  by the second quantization of $j_s$, namely
  $
    \Je_t\wick{A(\hat f_1)\cdots A(\hat f_n)}\Omega
 =
 \wick{\mathscr{A}^E(j_t f)\cdots \mathscr{A}^E(j_t f_n)}$ and $\Je_t\Omega=1$,
 where $\wick{\xi}$ denotes the Wick product of $\xi$.
 \bp{daiw}
 Let $F,G\in\hhh$ and $f_1,...,f_{n-1}\in L^\infty(\BR)$.
 For $S=t_0\leq t_1\leq \cdots\leq t_n=T$,
 \eqn
 &&
  (F,
 e^{-(t_1-t_0)\PF}
 (f_1\otimes1)
 \cdots (f_{n-1}\otimes1)
 e^{-(t_n-t_{n-1})\PF} G)_\hhh\\
 &&
=\E_\W\left[
           \lk
    \prod_{j=1}^{n-1} f_j(B_{t_j})\rk
e^{-\int_{S}^TV(B_s)ds}
 \E_{\mu_E}\left[\ov{\Je_{S}F(B_{S})}
 e^{
 -\I \alpha \int_{S}^{T}
   \mathscr{A}_s^E \cdot \D B_s}\Je_TG(B_T)
 \right]\right].
 \enn
\ep \proof See \cite{h4,h26}.
\qed

We are now ready
to state and prove the main theorem of this paper.
\bt{a5} Suppose that
Assumption (GS) and
\kak{kato}.
 Then there exists a subsequence $T'$ such that the
weak limit of $\mu^V_{T'}$ as $T'\rightarrow \infty$ exists. \et
 \proof
 By the Prohorov theorem, it is enough to show two
facts:\\
(1)
 $\d \lim_{\Lambda\rightarrow \infty}\sup_T
  \mu_T^V(|B_0|^2>\Lambda)=0$,\\
(2) for arbitrary $\epsilon>0$,
 $\d \lim_{\delta\downarrow 0} \sup_T\mu^V_T
\lk
\max_{\stackrel{|t-s|<\delta}{-T\leq s,t\leq T}}|B_t-B_s|>\epsilon\rk =0$.\\
  Using  Corollary \ref{exp}
we have
$$ \mu^V_T(|B_0|^2>\Lambda)
 = (\gr^T,
 (\chi_{\{|x|^2>\Lambda\}}\otimes 1) \gr^T)_\hhh,$$
 where $\chi_{D}$
  denotes the characteristic function on $D$.
   Using the fact that
 $\gr^T\rightarrow \gr$ strongly as $T\rightarrow \infty$ and
$\|\chi_{\{|x|^2>\Lambda\}}\gr\|_\hhh\rightarrow 0$ as $\Lambda\rightarrow
\infty$, we get (1). For (2), assume that $|t-s|$ is sufficiently
small.
 It is enough to show that
 \eq{3} \E_{\mu_T^V}[ |B_t-B_s|^{2n}]  \leq  |t-s|^n D
\en
 with some constant $D$ independent of $T$.
To apply Propositions \ref{exp} and \ref{daiw},
 we have to truncate the process $B_t^\mu$ as
 $$(B_t^\mu)_a(w):=\lkk\begin{array}{rc}
 -a,&B_t^\mu(w)\leq -a,\\
 B_t^\mu(w),&|B_t^\mu(w)|<a,\\
 a,&B_t^\mu(w)\geq a.
 \end{array}\right.$$
 and define
 the truncated multiplication operator $h_a^\mu$, $\mu=1,2,3$,  by
 $$h_a^\mu f (x)=
 \lkk\begin{array}{rc}
  -af(x),& x_\mu \leq -a,\\
 x_\mu f(x),&|x_\mu |<a,\\
 af(x),&x_\mu\geq a.
 \end{array}
 \right.
 $$ Note that
 \eq{div}
 |h_a(x)-h_a(y)|\leq |x-y|\quad x,y\in\BR,
 \en
   for all $a\geq 0$.
Since $h_a^\mu$ is bounded, we can see that
 \eqn
 && \EV{ |(B_t)_a-(B_s)_a|^{2n}}\\
  &&=\sum_{\nu=1}^3 \sum_{k=0}^{2n}
  \vvv {2n\\ k}(-1)^k
  \EV{
  (B_s^\nu)_a^k (B_t^\nu)_a ^{2n-k}}\\
 &&=\sum_{\nu=1}^3
 \sum_{k=0}^{2n}
 \vvv {2n\\ k}(-1)^k \lk
 \frac{}{}\!\! \lk
 h_a^\nu\otimes1\rk^k e^{-sH} \gr^T,
  e^{-(t-s){H}}\lk
  h_a^\nu\otimes 1\rk^{2n-k}  e^{+tH} \gr^T
   \rk _\hhh \\
 &&=\sum_{\nu=1}^3 \sum_{k=0}^{2n}\vvv{2n\\k}(-1)^k
 \E_\W \left[\frac{}{}   \lk \frac{}{}h_a^\nu(B_0)\rk ^k
 \lk \frac{}{} h_a^\nu(B_{t-s}) \rk^{2n-k}
  e^{-\int_0^{t-s} V(B_s) \D s}\right.\\&&
\hspace{3cm} \times \left. \frac{}{} \E_{\mu_E}\left[
 \ov{\Je_0e^{-sH} \gr^T(B_0)}  e^{-i\alpha \int_{-T}^T\mathscr{A}_s^E\cdot \D B_s}
 \Je_{t-s}e^{+tH}\gr^T(B_{t-s})\right]\right]
\\
&&=
 \E_\W\left[\frac{}{}  |h_a(B_0)-h_a(B_{t-s})| ^{2n}
  e^{-\int_0^{t-s} V(B_s) \D s}\right.\\
  &&
\hspace{3cm} \times\left.\frac{}{} \E_{\mu_E} \left[
 \ov{\Je_0e^{-sH}
 \gr^T(B_0)}
 e^{-i\alpha \int_{-T}^T \mathscr{A}_s^E\cdot \D B_s}
 \Je_{t-s}e^{+tH}\gr^T(B_{t-s})\right]\right]
\\
&&
 \leq
 \E_\W\left[\frac{}{}
  |B_0-B_{t-s}|^{4n}
  \|e^{+tH}
  \gr^T(B_{t-s})\|_\hhh^2 \right]^{1/2} \\
  && \hspace{3cm} \times
  \E_\W\left[ e^{-2\int_0^{t-s} V(B_s) ds}
\|e^{-sH}\gr^T(B_0)\|_\hhh ^2\right] ^\han \\
 && \leq C_V
 \|e^{+tH} \gr^T\|_\hhh\|e^{-sH}\gr^T\|_\hhh
 \E_{P_W^{0,0}}\left[\frac{}{}
 |B_0-B_{t-s}|^{4n}\right]^\han
\\
 && \leq |t-s|^{n} \sqrt{C_{4n}}
 C_V \|e^{+tH} \gr^T\|_\hhh\|e^{-sH}\gr^T\|_\hhh 
 \enn
where
we $e^{+tH}\gr^T$ is well defined for $t<T$,
 and we used Corollary \ref{exp} in the second equality,
 Proposition \ref{daiw} in the third equality, \kak{div}
 in the fifth
 inequality,
   $C_V:=\sup_{x\in\BR}
 \E_{P_W^x} [
 e^{-2\int_0^{t-s} V(B_r) dr}] <\infty$ and
    $C_{4n} $ is the constant such that
$$\E_{P_W^{0,0}}[ |B_s-B_t|^{4n}]=C_{4n}|t-s|^{2n}.$$
 Since
 $\|e^{-sH}\gr^T\|\rightarrow e^{-sE(H)}\|\gr\|$ and $\|e^{+tH}\gr^T\|\rightarrow e^{tE(H)}\|\gr\|$  as $T\rightarrow \infty$,
 we have
 $D:=\sup_T \sqrt{C_{4n}}
 C_V \|\gr^T\|_\hhh^2e^{(t-s)E(H)}<\infty$.
 Then we have
 $$\E_{\mu_T^V}[|(B_t)_a-(B_s)_a|^{2n}]\leq D|t-s|^n$$ uniformly in $a$. Since the left-hand side above monotonously increasing as $a\uparrow\infty$,
 the monotone convergence theorem yields \kak{3}.
Thus  (2) follows.
\qed
\begin{definition}\label{inf}
Let $V\in K$ and suppose Assumption (GS). Then the  weak limit of the measure
$\nu_{T'}^V$ on $\CCR$
is denoted by $\nu_\infty ^V$.
\end{definition}
Using the functional integration of $e^{-tH}$,
we can show
the Carmona type estimate \cite{ca},
namely    $\gr$ is spatially localized as follows:
if $V(x)=|x|^{2n}$, then $\|\gr(x)\|_\fff\leq C_1 e^{-C_2|x|^{n+1}}$,
and if $V(x)=-1/|x|$, then
$\|\gr(x)\|_\fff \leq C_3e^{-C_4|x|}$ for some constants $C_j$.
We have a corollary.
\bc{main2}
Assume that $\|\gr(x)\|_\fff\leq Ce^{-c|x|^\gamma}$ for some positive constants $C,c$ and $\gamma$. Then
\eq{loc}
\int_\CCR e^{c|B_0|^\gamma} \nu_\infty ^V(dw)<\infty.
\en
\ec
\proof
Let  $\rho_m(x)=\lkk \begin{array}{cc} e^{c|x|^\gamma},& e^{c|x|^\gamma}\leq m,\\
m, & e^{c|x|^\gamma}>m.\end{array}\right.$ Then $(\gr, (\rho_m\otimes 1)\gr)_\hhh=\int _\CCR \rho_m(B_0) \mu_\infty^V$ follows. By the limiting arguments as $m\rightarrow \infty$, we have \kak{loc}.
\qed

\subsection{Concluding remarks}
 In this paper we have given one example where we can
both make sense of the double stochastic integral and obtain the
infinite volume Gibbs measure by coupling Brownian motion to an
auxiliary Gaussian measure. The drawback of this particular example
is that the Gaussian space is infinite dimensional, and the
associated Hamiltonian along with the existence of its ground state
is non-trivial, and so we have to rely on a lot of technology. It is
conceivable that the same method should work in a much easier case,
namely when the auxiliary Gaussian process is just the stationary
one-dimensional (or $n$-dimensional) Ornstein-Uhlenbeck process.
However, when trying this approach one notices that on the way we
used a lot of special features of the Pauli-Fierz model and its
associated functional integral: for example the translation invariance
of the coupling ensures that the term arising from the diagonal does
not depend on $B_{t}$, which is a feature that cannot be
reproduced in finite dimension. So while we believe that a theory of
double stochastic integrals originating from the variance of a
Gaussian process could be developed, it is not altogether
straightforward and we leave it as a future project.

{\bf Acknowledgments} V.B.\ wishes to thank the University of Kyushu, Japan, where the present work was started, for kind hospitality.
F. H. wishes to thank the University of Warwick in UK, where this work was partially done, for kind hospitality.

{\footnotesize

\end{document}